\documentclass[%
 reprint,
 amsmath, assume,
 aps,
pra,
]{revtex4-2}

\usepackage[colorlinks,citecolor=blue,linkcolor=magenta,urlcolor=blue]{hyperref}  

\usepackage{braket}
\usepackage{xspace}
\usepackage{tikz,quantikz}
\usepackage{orcidlink}         
\usepackage{comment}
\usetikzlibrary{external}
\usepackage{amssymb}
\usepackage[T1]{fontenc}

\usepackage[ruled,vlined]{algorithm2e}
\usepackage[capitalise]{cleveref}	
\usepackage{import}
\usepackage{transparent}

\usepackage[mode=build]{standalone}

\usepackage{graphicx}
\usepackage{dcolumn}
\usepackage{bm}
\usepackage{physics}	
\usepackage{dsfont}

\usepackage{multirow}

\crefname{algorithm}{Algorithm}{Algorithms}
\crefname{appendix}{Appendix}{Appendices}
\crefname{section}{Section}{Sections}

\newcommand{\NN}{\mathcal{N}}

\makeatletter
\def\blfootnote{\xdef\@thefnmark{}\@footnotetext}
\makeatother

\begin{document}

\preprint{APS/123-QED}
\title{Graph state extraction from two-dimensional cluster states}
\author{ Julia Freund\orcidlink{0000-0001-5548-5007}$^{1}${{\large\color{blue}$^*$}}, Alexander Pirker\orcidlink{0000-0003-1260-1981}$^{1}$, Lina Vandr\'e\orcidlink{0000-0003-2753-6027}$^{2,3}$ and Wolfgang Dür\orcidlink{0000-0002-0234-7425}$^1${{\large\color{blue}$^*$}}}

 \affiliation{$^1$ Universität Innsbruck, Institut für Theoretische Physik, Technikerstraße 21a, 6020 Innsbruck, Austria}
 \affiliation{$^{2}$ Naturwissenschaftlich-Technische Fakult\"at, Universit\"at Siegen, Walter-Flex-Stra\ss e 3, 57068 Siegen, Germany}
 \affiliation{$^{3}$ Technische Universität Wien, Atominstitut, Vienna Center for Quantum Science and Technology, Stadionallee 2, 1020 Vienna, Austria}

\date{\today}

\begin{abstract}
We propose schemes to extract arbitrary graph states from two-dimensional cluster states by locally manipulating the qubits solely via single-qubit measurements. We introduce graph state manipulation tools that allow one to increase the local vertex degree and to merge subgraphs. We utilize these tools together with the previously introduced zipper scheme that generates multiple edges between distant vertices to extract the desired graph state from a two-dimensional cluster state. We show how to minimize overheads by avoiding multiple edges, and compare with a local manipulation strategy based on measurement-based quantum computation together with transport. These schemes have direct applications in entanglement-based quantum networks, sensor networks, and distributed quantum computing in general.
\end{abstract}

\maketitle

\blfootnote{{{\large\color{blue}$^*$}}Corresponding authors: \\
Julia Freund (\href{mailto:julia.freund@uibk.ac.at}{julia.freund@uibk.ac.at}), \\
Wolfgang D\"ur (\href{mailto:wolfgang.duer@uibk.ac.at}{wolfgang.duer@uibk.ac.at}).
}

\section{\label{sec:intro}Introduction}
The past few decades have been significantly influenced by advances in computation and the internet. However, we are now beginning to encounter the limits of these technologies. The transition from classical to quantum computing~\cite{nielsen_chuang_2010, HorodeckiActaPhysicaPolonicaA, Zoller2005} has the potential to overcome some of these limitations. Promising applications include quantum computing algorithms, such as Shor's~\cite{Shor1997primefactor,ShorAlgorithm1994}, Grover's~\cite{Groover1996} and Deutsch-Josza's~\cite{deutsch1989quantum} algorithm, as well as advancements in quantum sensing~\cite{Degen2017ReviewQSensing, Proctor2018, Guo2020, SekatskiPRR2020,Bugalho2025privaterobuststates}, secret sharing~\cite{ZhangPRA2024,Hillery99,Markham08} and the development of quantum networks~\cite{rohde2025quantum,vanmeterQuInternetArchitecture2022,  QuantumInternetWehner, QIalliance,EuroQCI}.

\begin{figure}
    \centering
    \includegraphics[width=0.7\linewidth]{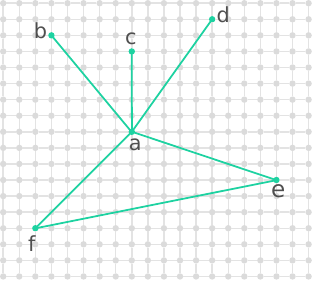}
    \caption{
    The family of graph states are multipartite entangled states which can be represented by graphs such as the green graph in the figure. A prominent graph state is the two-dimensional cluster state, whose underlying graph is a grid, as shown in gray. In this paper, we initially have two-dimensional cluster states and aim to extract general graph states from it.
    }
    \label{fig:setting}
\end{figure}

Multipartite quantum states are crucial for many of these applications, and among them Greenberger-Horne-Zeilinger (GHZ)~\cite{GreenbergerHornerZeilinger1989} states are frequently used. Graph states~\cite{heinEntanglementGraphstates}, a family of multipartite entangled states that have a graph representation as shown in \cref{fig:setting}, have proven to be particularly important. Graph states have applications in benchmarking quantum computing platforms~\cite{boensel2024generatingmultipartitenonlocalitybenchmark}, quantum networks~\cite{Pirker_2019,Pirker_2018,negrin2024efficientmultipartyentanglementdistribution}, and even serve as the foundation for universal quantum computation in measurement-based quantum computing (MBQC)~\cite{RaussendorfOneWayPRL2001}.

A resource for the mentioned applications is entanglement~\cite{Shahandeh2019}. Generating entangled multipartite graph states in general requires a high connectivity between the parties. However, this connectivity can be limited in many settings. In quantum computer platforms, gates are often only between nearest-neighbor qubits~\cite{BaeumerPRXQ2024,IBM,Starmon5Facts,Arute2019}. In quantum networks, with applications in key distribution \cite{Murta_2020_quantum_conference,ACKA,proietti2021experimental,expACKA} and quantum communication~\cite{Gisin2007,khatri2020principles}, a direct link is required through a quantum channel, where, again, the geometry of the network restricts the easily accessible entanglement structures. Distant parties of a quantum network can only perform operations on their local qubits, making global operations impossible. The availability of quantum channels restricts the possible interactions, and thus, the complexity~\cite{kumabe2024complexity} and limitations in the creation of graph states~\cite{Hahn2019, Azuma2021, Hansenne_2022, makuta2023no, vandre2024distinguishinggraphstatesproperties} are studied.

Instead of directly preparing the target state, it is sometimes beneficial and also easier to prepare a more general resource state and to manipulate this resource to extract the target entangled state. In fact, certain graph states provide a viable resource for this task, as supported by numerous related studies. Multiple works consider methods to extract quantum states from a pre-given graph state. The paradigmatic example of a graph state is the two-dimensional cluster state introduced in~\cite{BriegelPRL2001Cluster}, which can be used as a universal resource in measurement-based quantum computation (MBQC)~\cite{RaussendorfOneWayPRL2001,Mantri2017,RaussendorfPRA2003MBQCCluster}. A general characterization of universal resources for MBQC and criteria for universality have been established in~\cite{vanDenNestPRL_2006}. Universal MBQC enables the preparation of arbitrary quantum states on a subsystem. For communication purposes, Refs.~\cite{Hahn2019, freund2024flexiblequantumdatabus, szymanski2024usefulentanglementextractednoisy} discuss the extraction of single and multiple Bell states as target states from an arbitrary graph state, the two-dimensional cluster state, or families of noisy graph states, respectively. The extraction of GHZ states from a one-dimensional cluster state and a two-dimensional cluster state is demonstrated in Refs.~\cite{dejong2023extracting,frantzeskakis_extracting} and Refs.~\cite{basak2024improvedroutingmultipartyentanglement,mannalathPRAmultiparty}, respectively. Note, that the problem of extracting a graph state from another graph state with more vertices is also known as the vertex-minor problem \cite{Kim2024vertex, wagner1937vertexminoar}. The general problem is known to be NP-complete \cite{DahlbergNPcomplete}.

In this work, we study the following setting: The task at hand corresponds to flexibly extracting an arbitrary graph state from a two-dimensional cluster state shared between a selected subset of qubits. We allow only single-qubit operations, namely single-qubit measurements and single-qubit unitaries. \cref{fig:setting} shows the overall idea, where the two-dimensional cluster state corresponds to the gray grid, and the goal is to extract the green graph state between the qubits $a$ to $f$. We solve this task in a centralized and decentralized manner, where we find three schemes for extracting arbitrary graph states from a two-dimensional cluster state. We differentiate between centralized and decentralized approaches based on the spatial position in the two-dimensional cluster state where the desired graph state is constructed. Our main findings include the following:
\begin{enumerate}
\item[(i)] We introduce two graph state manipulation tools that allow one to merge vertices and expand the vertex degree in a two-dimensional cluster state.
\item[(ii)] A decentralized extraction approach that generates graph states by directly cutting them out of the two-dimensional cluster states. This involves vertex degree expansion, and generating edges between vertices.
\item[(iii)] A centralized approach that utilizes local state generation via MBQC, and transport to the desired target vertices.
\end{enumerate}

The decentralized approach uses the two graph state manipulation tools as elementary operations for generating the required vertex degree and merging subgraphs by measuring a single isolated vertex. Both the centralized and decentralized approach use the recently introduced zipper scheme~\cite{freund2024flexiblequantumdatabus} to establish edges between pairs of vertices of the target state. The zipper scheme is essential to avoid limitations due to crossing of edges. We concentrate on the manipulation of states under idealized conditions, and consider a noiseless scenario. Tools such as the recently introduced noisy stabilizer formalism~\cite{NSF_MorRuizPRA2023} could be used to study the influence of noise and imperfections on the proposed schemes for small target graph states.

Our main goal is the extraction of arbitrary graph states from a pre-given two-dimensional cluster state, where both the vertices of the target graph as well as the graph itself, which determines the desired entanglement structure, can be flexibly chosen. This is in contrast to Ref.~\cite{Hahn2019} and Ref.~\cite{freund2024flexiblequantumdatabus} where single Bell and GHZ states and multiple Bell states are focused, respectively. This work differs from Refs.~\cite{dejong2023extracting,frantzeskakis_extracting,basak2024improvedroutingmultipartyentanglement,mannalathPRAmultiparty} by considering arbitrary target graph states instead of GHZ states, and Refs.~\cite{dejong2023extracting,frantzeskakis_extracting} by having two-dimensional instead of one-dimensional cluster states as an initial graph state. Our work uses the zipper scheme as a method to establish edges between vertices in the two-dimensional cluster state, which is in contrast to extracting the desired graph state by cutting it out with Pauli $Z$ measurements, and it allows us to extract multiple graph states in parallel. Our work and the mentioned Refs.~\cite{Hahn2019,freund2024flexiblequantumdatabus,dejong2023extracting,frantzeskakis_extracting,basak2024improvedroutingmultipartyentanglement,mannalathPRAmultiparty,RaussendorfOneWayPRL2001, RaussendorfPRA2003MBQCCluster,Briegel2009} share the characteristic of being top-down approaches, which means that a specific multipartite resource state is available and transformed by operations to extract the desired state. The opposite of the top-down approach is the bottom-up approach that combines smaller states, very often Bell states, to the required target state, as demonstrated in Refs.~\cite{sen2023multipartite,ShimizuPRA2025,chelluri2024multipartite,Meignant_2019,Bugalho2023distributing,ViscardiIEEE2023,fan2024optimizeddistributionentanglementgraph}.

Our work is structured as follows: We briefly introduce graph states and their local manipulations in~\cref{sec:graph_states} together with the necessary graphical tools. In particular, in \cref{sec:manipulation_tools}, we introduce new graph state manipulation tools, most notable vertex degree expansion and merging of vertices, which we extensively use for our approaches to generate graph states from a two-dimensional cluster state. In~\cref{sec:generation_from_cluster}, we introduce the decentralized approach followed by the central generation approach, which we compare in~\cref{sec:comparison} with respect to the resources required to extract some prominent graph states. We discuss our work and give an outlook in~\cref{sec:discussion}.

\section{\label{sec:graph_states}Graph states and their manipulations}
In this section, we briefly review graph states and their manipulations, as these are the background for this work. We also review the existing zipper scheme. Moreover, we introduce two new graph state manipulation tools.

\subsection{Graph states}\label{sec:graph_states_sub}
A graph state~\cite{HeinPRA2004,heinEntanglementGraphstates} is a specific stabilizer state, which can be described by a classical graph $G=(V, E)$ consisting of a vertex set $V$ and an edge set $E$. The set of vertices $V$ corresponds to qubits, which are initialized in the positive eigenstate $\ket{+}$ of the Pauli $\sigma_{\mathrm{X}}$ operator. Edges of the set $E=\{\{a,b\} \mid a,b \in V, a \neq b\}$ denote entangling gates applied on the adjacent qubits.

With these ingredients, the graph state $\ket{G}$ corresponding to a graph $G = (V,E)$ is defined as
\begin{equation}
\ket{G}=\prod_{\{a, b\} \in E} \mathrm{CZ}_{a b}\ket{+}^{\otimes \vert V \vert},
\end{equation}
where a controlled-Z gate $\mathrm{CZ}_{a b}$ is applied on qubits $a$ and $b$ for all $\{a,b\} \in E$.

Examples of graphs are shown in \cref{fig:setting}. The graph state corresponding to the green graph is given by
\begin{align}
    \ket{G} = \mathrm{CZ}_{ab} \mathrm{CZ}_{ac} \mathrm{CZ}_{ad} \mathrm{CZ}_{ae} \mathrm{CZ}_{af} \mathrm{CZ}_{ef} \ket{+}^{\otimes 6}.
\end{align}

We define the neighborhood $\NN_a$ of a vertex $a \in V$ as the set of vertices adjacent to vertex $a$. That is for a graph $G=(V,E)$:
\begin{align}
    \NN_a \coloneqq \{ b \mid \{ a,b \} \in E \}.
\end{align}
Considering the green graph in \cref{fig:setting}, the neighborhood of vertex $e$ is given by $\NN_e = \{ a, f \}$. Graph states, corresponding to a graph $G = (V,E)$, can be manipulated by means of local unitaries and measurements, thereby generating other graph states on the remaining vertices, and these operations on graph states have a graphical representation \cite{Van_den_Nest_2004, heinEntanglementGraphstates}. Apart from creating a visual understanding of the gates, these graphical operations on the edge and vertex set are easier to perform than working with the density matrix formalism, which scales exponentially with the number of vertices $2^{\vert {V} \vert}$.
We review the effect of local Clifford and Pauli measurements on graph states.

\begin{figure}[t]
    \includegraphics[]{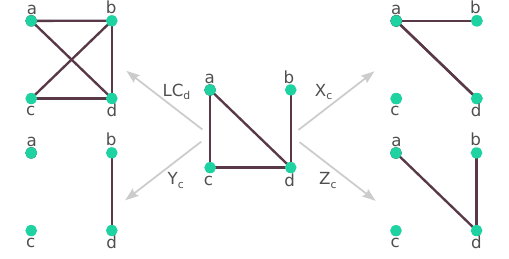}
    \caption{\label{fig:graphManipulations} The initial graph state, associated with the graph in the center, consists of four vertices (green dots) and four edges (gray lines). The states depicted around the central graph show the effects of local complementation and Pauli measurements.
    The up-left graphs show the effect of local complementation on vertex $d$. Vertex $d$ is adjacent to all other vertices. Because vertices $a$ and $c$ are adjacent in the initial graph, they get disconnected. Vertex $b$ is initially not adjacent to vertices $a$ and $c$ and therefore gets connected.
    If we measure qubit $c$ in the $Z$ basis, the resulting state is represented by the bottom-right graph: All edges containing $c$ are deleted. 
    To see the effect of measuring qubit $c$ in the $Y$ basis, we first apply a local complementation on qubit $c$ and then disconnect it from the graph. This is shown in the bottom-left graph.
    Finally, measuring qubit $c$ in the $X$ basis is represented by a local complementation on the adjacent vertex $d$, then applying the rules of the $Y$ measurement, and finally applying a local complementation on vertex $d$. The resulting graph is shown up-right.
    }
\end{figure}

The major graphical operation on a graph state $\ket{G}$ is local complementation at vertex $a$, which locally inverts the graph spanned by neighbors $\NN_{a}$ of $a$. This means that existing edges are removed if they have been there initially, or edges are added if they have not been there before. Note that the local complementation is its own inverse function, and it corresponds to the following operation:
\begin{align}
     LC_a = \sqrt{-\mathrm{i} X_a} \prod_{b \in \NN{a}} \sqrt{\mathrm{i} Z_b}, \label{eq:LCoperator}
\end{align}
where the definition of the square root of an operator is $\sqrt{\pm \mathrm{i} O} = e^{\pm \mathrm{i} \frac{\pi}{4} O}$. 
In the upper-left corner of~\cref{fig:graphManipulations} we show the effect of a local complementation on vertex $d$ of the central graph state. It was shown in Ref.~\cite{Van_den_Nest_2004}, that local Clifford operations \cite{nielsen_chuang_2010} on graph states can be expressed as a series of local complementation.

Also, single-qubit Pauli measurements have a graphical understanding. We show examples in \cref{fig:graphManipulations}. The most straightforward operation is the Pauli $Z$ measurement on vertex $v$ for which one disconnects $v$ from the graph by deleting all edges connected to it. A Pauli $Y$ measurement translates to performing first a local complementation on the vertex, followed by a $Z$ measurement. A Pauli $X$ measurement requires performing local complementation on a neighbor of the vertex, followed by measuring the vertex in the $Y$ basis, and finally, another local complementation of the same chosen neighbor. 
Note that in order to get the precise post-measurement states, local unitaries that depend on the measurement outcomes have to be applied to the presented graph states \cite{heinEntanglementGraphstates}. For simplicity, we leave the corrections to the end.

Our fundamental resource in this work is the two-dimensional cluster state, known from measurement-based quantum computing~\cite{RaussendorfOneWayPRL2001}. The cluster state is defined on a graph that is a two-dimensional, rectangular, plain grid, where the vertices are located at the intersections of the grid lines. In~\cref{fig:setting,fig:cluster_zipper} we show a part of a two-dimensional grid graph in gray.

\subsection{Zipper scheme} \label{sec:zipper}
\begin{figure}
    \centering
    \includegraphics[width=\linewidth]{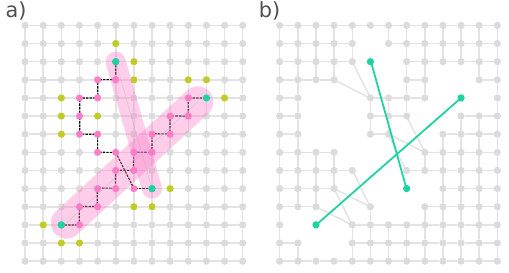}
    \caption{The two-dimensional cluster states correspond to a two-dimensional grid graph, shown in gray. a) The zipper scheme connects pairs of distant green qubits in the cluster state by measuring the pink qubits at a staircase-shaped path corresponding to the black dashed line. We illustrate the use of the zipper scheme and its iterative application by the boxes highlighted in pink (although the actual path may not be directly along that path). b) The post-measurement state is a tensor product of the two Bell pairs in green, and the graph state corresponding to the remaining grid graph. The zipper scheme mostly restores the underlying two-dimensional cluster state. Qubits represented by vertices adjacent to the end or turning points get measured in Pauli $Z$ basis, those are shown in bright yellow.} 
    \label{fig:cluster_zipper}
\end{figure}

In this section, we examine the zipper scheme~\cite{freund2024flexiblequantumdatabus}, which connects two distant vertices in a two-dimensional cluster state. This process generates a Bell state or an edge between two vertices while partly preserving the structure of the underlying grid graph and partly generating a structure with a similar connectivity, as illustrated in~\cref{fig:cluster_zipper}. For connecting two vertices, we first choose a path between them. Two examples of paths between distant (green) vertices are drawn as black dashed lines in~\cref{fig:cluster_zipper}\textcolor{magenta}{a}. It is an advantage to choose staircase-shaped paths, as shown in one of the examples from \cref{fig:cluster_zipper}\textcolor{magenta}{a}. However, in some cases there exists no direct staircase-shaped path between the vertices, in such cases we choose a path with merged staircase-shaped and straight segments. Note that straight-line segments require removal of all neighbors that surround the path, which cuts a hole in the remaining cluster state. All qubits associated with the vertices on the path are measured in Pauli $X$ basis, and we color them pink in~\cref{fig:cluster_zipper}. According to the measurement rules, introduced in \cref{sec:graph_states_sub}, this introduces edges between the chosen vertices, but also between them and other vertices in the grid. After measuring the qubits associated to vertices along the path, the chosen vertices are connected to vertices on both sides of the straight paths, as well as to most vertices adjacent to the chosen ones. Those vertices are colored in yellow in the figure. In order to isolate the chosen pair of vertices, we measure the qubits associated to the yellow vertices in the Pauli $Z$ basis. Ref.~\cite{freund2024flexiblequantumdatabus} discusses in detail, which of the qubits have to be measured in $Z$ basis in order to ensure that the connected Bell pairs get isolated. The resulting graphs are shown in~\cref{fig:cluster_zipper}\textcolor{magenta}{b}.

Note that the structure of the grid remains almost the same after the measurement process. Especially, vertices on both sides of the staircase-shaped paths get connected by new edges. The grid along straight paths, is deleted. See \cref{fig:cluster_zipper}\textcolor{magenta}{b} for an illustration. The graph associated to the post-measurement state can be understood intuitively by considering the effects of Pauli $X$ and $Z$ measurements: In the case of a $Z$ measurement, we delete a vertex and the adjacent edges, which reduces the connectivity of the graph, while for a $X$ measurement, we additionally create new edges. A detailed analysis can be found in Ref.~\cite{freund2024flexiblequantumdatabus}. The regeneration of the underlying grid allows us to cross measurement paths, which enables us to extract non-plane or multiple graph states from a single cluster state.

\subsection{Additional manipulation tools}\label{sec:manipulation_tools}
In this subsection, we introduce the two new tools that we utilize in this work to extract graph states from a two-dimensional cluster state. The first is a strategy to merge two connected vertices in an existing graph state with only local measurements. The second tool is the vertex degree expansion, which is used to extend the neighborhood of a vertex in a two-dimensional cluster state.

\subsubsection{Merging of two subgraphs}\label{sec:merging_sub_graphs}

\begin{figure}
    \centering
    \includegraphics[width=\columnwidth]{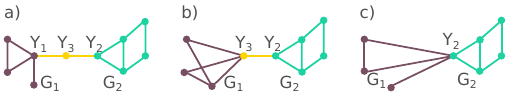}
    \caption{Merging of two subgraphs: a) The yellow path of edges is necessary to merge the two subgraphs $G_1$ (brown) and $G_2$ (green). b) The first step is to measure the qubit associated to the vertex labeled $Y_1$ in the Y basis, which inverts the neighborhood of $Y_1$ and connects it to $Y_3$. c) A $Y$ measurement of $Y_3$ merges $G_1$ and $G_2$ at vertex $Y_2$, while restoring the original neighborhood of vertex $Y_1$. The roles of $Y_1$ and $Y_2$ can be interchanged, leading to a merging at $Y_1$. }
    \label{fig:general_merging}
\end{figure}

We determine the necessary structure to merge the vertices of two adjacent subgraphs $G_1$ and $G_2$ embedded in the graph $G$, as shown in~\cref{fig:general_merging}. The aim is to find a solution that preserves the overall structure of $G_1$ and $G_2$ and merges them at one common vertex using only local measurements and unitaries. The subgraphs must be connected by a path of edges, which are not part of the graphs $G_1$ and $G_2$. Since a path graph can be shortened by applying Pauli $Y$ measurements, we assume that the path consists only of two edges in the following. We sketch the situation in~\cref{fig:general_merging}: the subgraphs $G_1$ and $G_2$ are colored brown and green, respectively, and the two edges connecting the two subgraphs are colored in yellow. 

The first Pauli $Y$ measurement is applied to the qubit associated with the vertex labeled $Y_1$, which is the vertex of graph $G_1$ that is connected to the vertex $Y_2$ in $G_2$ by a two-edge path. The second Pauli $Y$ measurement is applied to the qubit associated with the vertex labeled $Y_3$, which is the middle vertex of the connecting path.
The reason for having two Pauli $Y$ measurements is that the first measurement on $Y_1$ inverts the neighborhood structure of the subgraph $G_1$ and connects all neighbors to the middle vertex of the connecting path. This inversion of the neighborhood structure is compensated again by the measurement of qubit $Y_3$, and as such, it restores the subgraph structure and finalizes the merging procedure. We show the measurement pattern in~\cref{fig:general_merging}, which allows, due to symmetry, to exchange the roles of $G_1$ and $G_2$. We note that the structure of the subgraphs may allow having only a single measurement, for example, when $G_1$ and $G_2$ are linear cluster states. Note that in the two-dimensional cluster state, one may meet the criteria for merging two subgraphs $G_1$ and $G_2$ by isolating a path that connects $G_1$ and $G_2$ via Pauli $Z$ measurements together with Pauli $Y$ measurements to merge the linear cluster state and to obtain the setting in~\cref{fig:general_merging}\textcolor{magenta}{a}. 

This method does not allow any edge between the vertex $Y_1$ and any vertex associated with $G_2$, because the $Y$ measurement of $Y_1$ establishes edges of all vertices in $G_1$ that neighbor $Y_1$ and the neighbors that belong to $G_2$. The same condition holds for vertex $Y_3$ and the vertices in $G_1$.

\subsubsection{Vertex degree expansion}\label{sec:vertex_degree_expansion}
\begin{figure}
    \centering
    \includegraphics[width=\columnwidth]{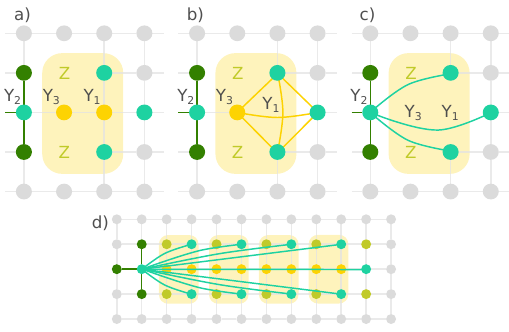}
    \caption{Vertex degree expansion: a-c) The vertex degree expansion increases the degree of the bright-green vertex $Y_2$ from $d=4$ to $d=6$, where the dark green qubits and edges denote the already existing neighbors. a) Two Pauli $Z$ measurements, labeled as bright-yellow $Z$, isolate the edge to be expanded from the grid. Measurement of the yellow qubit $Y_1$, seen in b), and $Y_3$, seen in c), expand the degree of the vertex $Y_2$. d) A $n_{\text{exp}}$ times iterative application of the vertex degree expansion along the horizontal direction enhances the degree of vertex $a$ by $2\times n_{\text{exp}}$. }
    \label{fig:degree_expansion}
\end{figure}

The vertex degree in a two-dimensional cluster state limits the number of edges of a vertex to four. However, for general graphs, the degree is only restricted by the number of vertices. For example vertex $a$ of the graph in~\cref{fig:setting} has a degree of five. Therefore, we introduce the following new graph state manipulation tool that allows us to expand the degree of a vertex of the two-dimensional cluster state by applying local Pauli measurements. We sketch our method in~\cref{fig:degree_expansion}\textcolor{magenta}{a-c}. As demonstrated, this method allows us to expand the degree of a vertex of the grid graph from $d=4$ to $d=6$. First, we isolate the vertex attached to the edge we want to expand, by measuring the qubits associated to the neighboring vertices in the Pauli $Z$ basis; see~\cref{fig:degree_expansion}\textcolor{magenta}{a}. Then, the vertex expansion is completed by measuring qubit $Y_1$ followed by qubit $Y_3$ in the $Y$ basis; which we show in~\cref{fig:degree_expansion}\textcolor{magenta}{b} and \textcolor{magenta}{c}. This method increases the vertex degree by two while requiring $n=4$ measurements embedded in a $2\times 3$ grid. We introduce yellow boxes, as one sees in~\cref{fig:degree_expansion}, to indicate the use of the vertex degree expansion. The vertex degree expansion can be seen as a specific instance of merging two subgraphs. 

There are several ways to use the vertex degree expansion method iteratively, and here we discuss only the unidirectional iteration. We refer to~\cref{app:vertex_iteratively} for other ways to apply the vertex degree expansion iteratively. 

The unidirectional expansion iterates the vertex expansion method along the horizontal or vertical direction in the two-dimensional grid graph respectively.~\cref{fig:degree_expansion}\textcolor{magenta}{d} shows the iterative horizontal vertex degree expansion. Applying the vertex degree expansion $n_{\text{exp}}$ times requires $4\times n_{\text{exp}}$ measurements and increases the degree by $2\times n_{\text{exp}}$. This method scales linearly with the number of times the vertex degree expansion is applied. The expanded vertices are on opposite sides of the expansion area, which may not be ideal for every case. Thus, an alternative may be to apply the method at all four edges of the initial vertex. We refer to~\cref{app:vertex_iteratively} for an alternative iterative application of the method.

\section{Graph state extraction approaches}\label{sec:generation_from_cluster}

In this section, we focus on generating multipartite graph states within a two-dimensional cluster state using only local operations and Pauli measurements. We introduce a decentralized and centralized approach for graph state extraction. Within the decentralized approach, we find two schemes to extract the graph state. Note that we assume that the vertices in the graph are fixed; thus, we cannot substitute them with vertices in the neighborhood to create an equivalent graph state of the same structure.

\subsection{Decentralized approach} \label{sec:delocalized_approach}
The idea of the decentralized approach is to extract the desired graph state in a distributed fashion from the two-dimensional cluster state. We present the local vertex degree expansion scheme and the optimized vertex degree expansion scheme as instances of the decentralized approach. Both schemes utilize the graph manipulation tools we introduced in~\cref{sec:manipulation_tools}. 

\subsubsection{Local vertex degree expansion}
\begin{figure}
    \centering
    \includegraphics[width=\columnwidth]{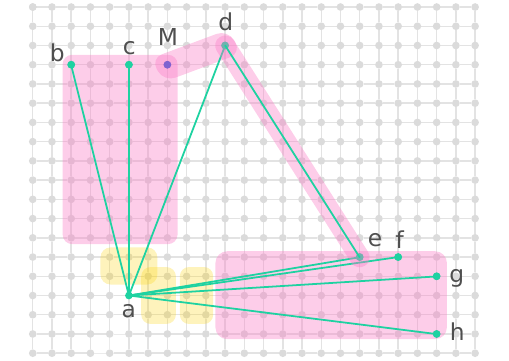}
    \caption{Local vertex degree expansion: The aim is to extract the state corresponding to the green graph from the two-dimensional cluster state. We first expand the vertex degree, if needed. In this example, the degree of vertex $a$ has to be increased to 7, which is done by using the vertices in the yellow boxes. Then, the zipper scheme establishes the edges between the vertices as indicated by the pink boxes. The purple vertex $M$ merges the subgraphs attached to the qubits $a$ and $d$. }
    \label{fig:local_extraction}
\end{figure}

The local vertex degree expansion is the first of the two schemes of the decentralized approach. As illustrated in~\cref{fig:local_extraction}, this scheme aims to extract the green graph from the gray grid graph. The main idea of this scheme is to directly use the vertex degree expansion of~\cref{sec:vertex_degree_expansion} at each vertex, which we indicate by the
yellow boxes in~\cref{fig:local_extraction}. When all vertices have the required degree, we use the zipper scheme of~\cref{sec:zipper} to establish the required edges between the vertices, which we illustrate by the pink boxes in~\cref{fig:local_extraction}. If necessary, the merging tool of~\cref{sec:merging_sub_graphs} allows us to connect subgraphs, the purple qubit $M$ in~\cref{fig:local_extraction} merges the subgraphs attached to the qubits $a$ and $d$.
  
The total number of measurements needed for this scheme depends on the individual vertex degree expansion and the contributions from the zipper scheme measurements. The vertex degree expansion requires $4\times n_{\text{exp}}$ measurements to expand the degree by $2\times n_{\text{exp}}$ as discussed in~\cref{sec:vertex_degree_expansion}. To establish $n_{\text{e}}$ edges by the zipper scheme, it approximately requires $n_{\text{e}} \times N$ measurements, where $N$ is the average distance in a $N\times N$ cluster state. 

Large vertex degrees limit the number of possible neighbor configurations this scheme can provide because the vertex degree expansion requires a space of $4\times n_{\text{exp}}$ measurements. Moreover, we apply the zipper scheme individually for each edge in the target graph, which constrains the achievable graph state configurations further. For example, the four green vertices in the lower right and upper left corner of~\cref{fig:local_extraction} require individual measurement paths for each vertex, although they are in the same region of the cluster state. We show how to overcome those issues with the scheme we introduce next.

\subsubsection{Optimized vertex degree expansion} 
With the scheme presented in this section, we want to solve the primary limitation of the previous method. That is, we address the significant requirement of resources due to high vertex degrees and the measurement paths necessary to connect the individual vertices. The main idea of this approach is to improve the state extraction strategy by reducing the number of measurement paths, thereby also reducing the number of measured qubits, due to clever combination of the measurement paths and the merging tool.

\begin{figure}
    \centering
    \includegraphics[width=0.9\linewidth]{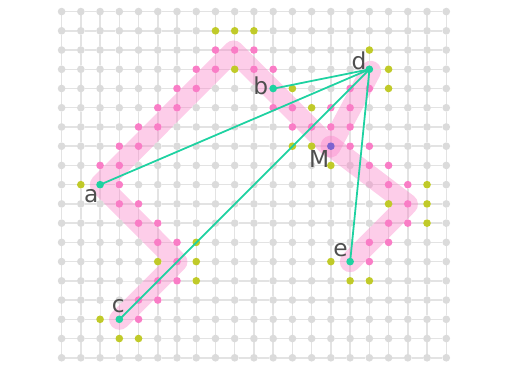}
    \caption{Path optimization: The zipper scheme, indicated by pink boxes, and the merging tool applied to the purple qubit $M$ facilitate the generation of a GHZ state among the green qubits. This approach optimizes the local vertex degree expansion by combining the zipper scheme with the merging tool. In this case the purple vertex $M$ merges vertex $d$ and the pre-established GHZ state. }
    \label{fig:collecting_GHZ}
\end{figure}

We aim to find an optimized measurement path together with the positions of junction points, allowing us to extract the graph state from the two-dimensional cluster state with a significantly reduced number of measurements.~\cref{fig:collecting_GHZ} shows the optimized vertex degree expansion for extracting the green GHZ state that corresponds to a star graph with central vertex, $d$, which has edges to all other vertices, $a-c$ and $e$. Please refer to the caption for details. It may be necessary to use the merging tool of~\cref{sec:merging_sub_graphs} at junction points e.g., purple qubit $M$ in~\cref{fig:collecting_GHZ}, to establish the desired edges. We discuss the cases when it requires the merging in~\cref{app:optimized_degree_expansion} in detail. Note that the merging tool requires measuring adjacent qubits at the junction points, which we discuss for a vertex with a degree of four in~\cref{app:optimized_degree_expansion}.

In~\cref{alg:GHZ_mesurement}, we describe an algorithm for finding an optimized measurement path to extract a GHZ state. At its core, the algorithm iteratively searches among all vertices in the GHZ state for the closest vertex to the measurement pattern $P$ until all vertices are connected. We apply Dijkstra's algorithm~\cite{DIJKSTRA1959} with the Manhattan metric~\cite{Riesz1910,minkowski1910} between each point of the measurement pattern $P$ and the remaining vertices to find the next vertex $n$ to connect and the position of the junction vertex $j$. Then we connect these two points, $n$ and $j$, with the zipper scheme such that there are at most two staircases beginning from $n$ and $j$, for which we present the algorithm in~\cref{app:one_turn_zipper_algorithm}. The measurement pattern we present in~\cref{fig:collecting_GHZ} corresponds to the outcome of applying~\cref{alg:GHZ_mesurement} to the green vertices. Note that we add the junction point according to the rules we discuss in~\cref{app:optimized_degree_expansion}. It is important to note that this algorithm can be applied iteratively for each vertex to determine a measurement pattern that extracts an arbitrary graph state from the two-dimensional cluster state. This can be easily seen as a result of the following observation: For a graph state $G$, one can start by generating the adjacency of a vertex $v$ by performing~\cref{alg:GHZ_mesurement} for the GHZ state that has its center located at $v$ and connects to all neighbors of $v$. Subsequently, one continues performing that procedure on the remaining vertices of $G$, whereas one excludes already established edges.

\begin{algorithm}
    \caption{Measurement path with junctions for GHZ state}
    \label{alg:GHZ_mesurement}
    \KwIn{Array of vertices to be in GHZ state $S$}
    \KwOut{Cluster state measurement pattern array $P$ }
    $P \gets 0$ ; \\
    $P(x,y) \gets 1 $ \tcp*{mark first vertex at $x$,$y$}
    $S \gets S \setminus \{S(0)\}$ \tcp*{remove first vertex from list}
    \While{$S \neq \emptyset$}{
    \tcc{returns next qubit $n$ and junction $j$}
    $n$, $j \gets$ Dijkstra($P$,$S$) ; \\
    \tcc{get zipper scheme measurement pattern}
    $P \gets$ one\_turn\_zipper($n$,$j$,$P$) ;\\
    $S \gets S \setminus \{n\}$ \tcp*{remove vertex $n$ from list}
    }
    \Return $P$ ;
\end{algorithm}

\begin{figure}
    \centering
    \includegraphics[width=\linewidth]{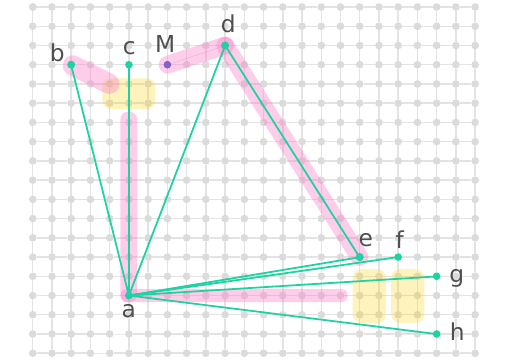}
    \caption{Optimized vertex degree expansion: The vertex expansion (yellow boxes), the zipper scheme (pink boxes), and the merging tool enable us to extract the state corresponding to the green graph from the two-dimensional grid graph. We place the yellow vertex expansion regions in areas where vertices cluster that should be connected to the same vertex. This approach reduces the number of times we need to employ the zipper scheme to just once. The purple qubit $M$ merges the subgraphs between $d$ and $e$.}
    \label{fig:distributed_extraction}
\end{figure}

The optimized vertex degree expansion scheme provides more flexibility in circumventing neighborhood constraints because it distributes the vertex expansion area within the cluster state and reduces the number of parallel measurement paths to the same destination. The optimized vertex degree expansion can be improved by using just a single edge of a vertex to reach an available space for the vertex degree expansion. In~\cref{app:optimized_degree_expansion}, we sketch an algorithm for finding free spaces that might be used for a delayed vertex degree expansion. In~\cref{fig:distributed_extraction}, we show how the optimized vertex degree expansion improves the measurement scheme of~\cref{fig:local_extraction}. The optimized vertex degree expansion needs fewer measurements because it uses the zipper scheme (pink boxes) only once to connect the (yellow) vertex degree expansion areas in the left and bottom right areas.

\subsection{Central generation}\label{sec:central_generation}
\begin{figure}
    \centering
    \includegraphics[width=\linewidth]{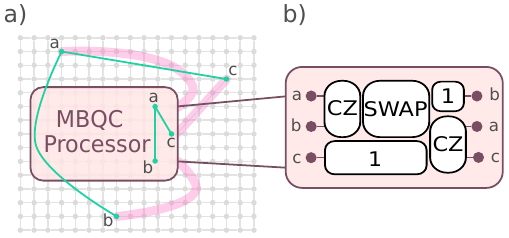}
    \caption{a) A central MBQC processor, embedded in the two-dimensional cluster state, generates the graph state $\ket{G}$ as illustrated by the green three-qubit GHZ state. Then, the output qubits of the computation are connected (green edges) and transported (pink boxes) to the vertices that are supposed to share the graph state. b) The MBQC algorithm utilizes $\operatorname{CZ}$ gates to establish edges together with $\operatorname{SWAP}$ gates to account for different scenarios. 
    }
    \label{fig:central_MBQC}
\end{figure}

In this section, we introduce an alternative method to generate graph states from a two-dimensional cluster state. This approach aims to generate the desired graph state in a central region using an MBQC processor, which is an area in the two-dimensional cluster state that prepares the state via measurements. MBQC allows universal quantum computing~\cite{RaussendorfPRA2003MBQCCluster}, which includes generating any arbitrary graph state.~\cref{fig:central_MBQC}\textcolor{magenta}{a} shows the main idea: In the center is the MBQC processor that generates a graph state as illustrated by the green three-qubit GHZ state. After the MBQC computation prompts the graph state, the zipper scheme and the merging procedure as described in~\cref{sec:zipper} and \cref{sec:merging_sub_graphs}, respectively, connect the output vertices with the desired distributed vertices i.e., transport the graph state to the desired target locations.

We propose the following algorithm for the MBQC computation, which idea we sketch in~\cref{fig:central_MBQC}\textcolor{magenta}{b}. An MBQC processor uses single-qubit measurements in the surrounding two-dimensional cluster state to implement gates such as a $\operatorname{CZ}$ or a $\operatorname{SWAP}$ gate, see Ref.~\cite{RaussendorfOneWayPRL2001,RaussendorfPRA2003MBQCCluster} for the required measurement pattern. 

In~\cref{fig:central_MBQC}\textcolor{magenta}{b} we illustrate an MBQC algorithm to generate a three-qubit GHZ state. In the first step, a $\operatorname{CZ}$ gate establishes the edge between the vertices $a$ and $b$. Then, the $\operatorname{SWAP}$ gate between the first two vertices $a$ and $b$ changes their spatial position, allowing us to establish the edge between $a$ and $c$ with another $\operatorname{CZ}$ gate.

For generating an arbitrary $N$-qubit graph state, we propose swapping the first vertex step by step from the first to the last position, thereby establishing edges if required. We continue in that spirit for all the remaining vertices; the second vertex is swapped through until the second last position, and so on. In the worst case, if all vertices need to be swapped, this procedure requires a two-dimensional cluster state of size $\mathcal{O}(N \times N^2)$ for the MBQC processor. This scheme is beneficial when the resulting graph has low vertex degrees such that only a few swaps are necessary.

\section{Comparison of the approaches} \label{sec:comparison}
In this section, we compare the performance of the local vertex degree expansion (LVDE), the optimized vertex degree expansion (OVDE), and the central generation (CG) of~\cref{sec:delocalized_approach} for extracting different types of graph states. Moreover, we comment on the features of the presented approaches.

\begin{figure}[h!]
    \centering
    \begin{tabular}{|c|c|c|c|}
        \hline
        extracted state& method & \begin{tabular}{@{}c@{}}vertex  \\ deg.\ prep.\end{tabular} & connect edges \\
        \hline
        \multirow{3}{*}{ 
            \parbox[c][3cm][t]{2.2cm}{ 
                \centering
                \vspace{-0.2cm}
                \includegraphics[width=0.9\linewidth]{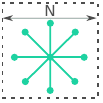} \\ 
                $n$-qubit GHZ\\
                $n_e = (n-1)$
            }
        } & \parbox[c][1cm][c]{1cm}{ LVDE } & \parbox[c][1cm][c]{1cm}{ $2\times n_e$ } & \parbox[c][1cm][c]{2.2cm}{ $n_e\times N$ } \\
        \cline{2-4}  
        & \parbox[c][1cm][c]{1cm}{ OVDE } & \parbox[c][1cm][c]{1cm}{ -- } & \parbox[c][1cm][c]{2.2cm}{ $n_e\times N$ } \\
        \cline{2-4}
        & \parbox[c][1cm][c]{1cm}{ CG } & \parbox[c][1cm][c]{1cm}{ $n_e\times n_e$ } & \parbox[c][1cm][c]{2.2cm}{ $n_e \times N$ } \\
        \hline
        \multirow{3}{*}{ 
            \parbox[c][3.6cm][t]{2.2cm}{ 
                \centering
                \vspace{-0.2cm}
                \includegraphics[width=0.9\linewidth]{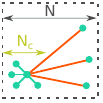} \\ 
                $n$-qubit GHZ \\
                $n_e = (n-1)$\\
                \textcolor[HTML]{ff5402}{$n_l$} \textsuperscript{*}, \textcolor[HTML]{c0ca29}{$N_c<< N$}
            }
        } & \parbox[c][1.2cm][c]{1cm}{ LVDE } & \parbox[c][1cm][c]{1cm}{ $2\times n_e$ } & \parbox[c][1cm][c]{3cm}{ $(n_e-1)\times N_c+N$ } \\
        \cline{2-4}  
        & \parbox[c][1.2cm][c]{1cm}{ OVDE } & \parbox[c][1cm][c]{1cm}{ -- } & \parbox[c][1cm][c]{3cm}{ $n_e\times N_c+N$ } \\
        \cline{2-4}
        & \parbox[c][1.2cm][c]{1cm}{ CG } & \parbox[c][1cm][c]{1cm}{ $n_e\times n_e$ } & \parbox[c][1cm][c]{3cm}{ $n_e \times N_c + N$ } \\
        \hline
        \multirow{3}{*}{ 
            \parbox[c][3cm][t]{2.2cm}{ 
                \centering
                \vspace{-0.2cm}
                \includegraphics[width=0.9\linewidth]{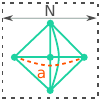} \\ 
                $K_n\setminus\{a\}$ \textsuperscript{**} \\
                $n_e = (n^2 -1)$
            }
        } & \parbox[c][1cm][c]{1cm}{ LVDE } & \parbox[c][1cm][c]{1.2cm}{ $n_e \times n_e^2$ } & \parbox[c][1cm][c]{3cm}{ $n_e\times n_e \times N$ } \\
        \cline{2-4}  
        & \parbox[c][1cm][c]{1cm}{ OVDE } & \parbox[c][1cm][c]{1.2cm}{ -- } & \parbox[c][1cm][c]{3cm}{ $n_e \times n_e \times  N$ } \\
        \cline{2-4}
        & \parbox[c][1cm][c]{1cm}{ CG } & \parbox[c][1cm][c]{1.2cm}{ $n_e\times n_e$ } & \parbox[c][1cm][c]{3cm}{ $n_e \times N$ } \\
        \hline
        \multirow{3}{*}{ 
            \parbox[c][3cm][t]{2.2cm}{ 
                \centering
                \vspace{-0.2cm}
                \includegraphics[width=0.9\linewidth]{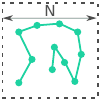} \\ 
                1D cluster \\
                $n_e = n-1$
            }
        } & \parbox[c][1cm][c]{1.2cm}{ LVDE } & \parbox[c][1cm][c]{1.2cm}{ -- } & \parbox[c][1cm][c]{3cm}{ $N$ } \\
        \cline{2-4}  
        & \parbox[c][1cm][c]{1.2cm}{ OVDE } & \parbox[c][1cm][c]{1.2cm}{ -- } & \parbox[c][1cm][c]{3cm}{ $N$ } \\
        \cline{2-4}
        & \parbox[c][1cm][c]{1.2cm}{ CG } & \parbox[c][1cm][c]{1.2cm}{ $n_e \times 1$ } & \parbox[c][1cm][c]{3cm}{ $n_e \times N$ } \\
        \hline
        \multirow{3}{*}{ 
            \parbox[c][3.6cm][t]{2.2cm}{ 
                \centering
                \vspace{-0.2cm}
                \includegraphics[width=0.9\linewidth]{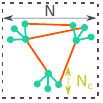} \\ 
                arbitrary \\
                $n_e$\textsuperscript{***}, \textcolor[HTML]{ff5402}{$n_l$}\textsuperscript{*}, \\
                \textcolor[HTML]{c0ca29}{$N_c<< N$}
            }
        } & \parbox[c][1.2cm][c]{1.2cm}{ LVDE } & \parbox[c][1cm][c]{1.3cm}{ $2\times (n_e-n_l)$ } & \parbox[c][1cm][c]{3cm}{ $(n_e-n_l)\times N_c+n_l\times N$ } \\
        \cline{2-4}  
        & \parbox[c][1.2cm][c]{1.2cm}{ OVDE } & \parbox[c][1cm][c]{1.3cm}{ -- } & \parbox[c][1cm][c]{3cm}{ $(n_e-n_l)\times N_c+n_l\times N$ } \\
        \cline{2-4}
        & \parbox[c][1.2cm][c]{1.2cm}{ CG } & \parbox[c][1cm][c]{1.3cm}{$(n_e-n_l)^2$} & \parbox[c][1cm][c]{3cm}{ $n_e \times N$ } \\
        \hline
    \end{tabular}

    \caption{The table compares the performance of three schemes we introduce in~\cref{sec:delocalized_approach,sec:central_generation} for extracting specific graph states from a $N\times N$ two-dimensional cluster state. The first column displays the graph state to be extracted, which has $n$ vertices and $n_e$ edges. We use the following abbreviations: LVDE for local vertex degree expansion, OVDE for optimized vertex degree expansion, and CG for central generation.}

    \vspace{3pt} 
    \parbox{\linewidth}{\raggedright \small \textsuperscript{*} \textcolor{red}{$n_l$} (red) long-range edges of $\mathcal{O}(N)$ }
    \parbox{\linewidth}{\raggedright \small \textsuperscript{**} fully connected graph missing the red edge $a$}
    \parbox{\linewidth}{\raggedright \small \textsuperscript{***} number of edges $n_e$ independent of number of vertices $n$}
    \label{tab:comparision_schemes}
\end{figure}

The three schemes presented in this work have distinct features and advantages, and their performance highly depends on the specific graph state to be extracted. Thus, we discuss the costs of extracting specific graph states with $n$ vertices and $n_e$ edges from the $N\times N$ two-dimensional cluster state in~\cref{tab:comparision_schemes}. In the first column of~\cref{tab:comparision_schemes}, we show the graph states we consider, which we describe in the following paragraphs. The second, third, and fourth columns represent the methods and the order of the required qubits to establish the vertex degree and edges.

The first state is a $n$-qubit GHZ states with $n_e$ long-range edges of length $\mathcal{O}(N)$. For that case, the OVDE is best suited because it only demands the measurements for establishing the edges, and then the LVDE and CG need linear and quadratic edges to prepare the vertex degree. 

The second state is a $n$-qubit GHZ with $n_e$ edges from which $n_l$ long-range edges of length $\mathcal{O}(N)$ and $n_e-n_l$ short-range edges of the length of $\mathcal{O}(N_c)$ with $N_c << N$ are demanded. This example differs from the previous merely by having different length scales in connecting the edges, and thus, the best suitable is again the OVDE followed by LVDE and CG.

The third graph state is a fully connected state that misses the (dashed, red) edge $a$ and has $n_e = n^2 -1$ long-range edges of length $\mathcal{O}(N)$. For this case, the CG and the OVDE are equally good since both scales quadratically in $n_e$ in total, whereas the LVDE scales with $n_e^3$.

The fourth graph state is a linear cluster state of total length $N$ comprised of $n_e$ edges. For that state, the LVDE and OVDE become the same method, which suits best of the extraction by scaling constantly by the length $N$ of the linear cluster state. For example, the CG scales linearly regarding the number of edges.

The last state is an arbitrary graph state with $n_l$ (red) long-range edges and $n_e-n_l$ edges of length $N_c$, as shown in the sketch. The most suitable method is the OVDE expansion, which scales linearly in the number of edges. Also, the LVDE scales linearly with the number of edges but requires additionally a linear number of measurements to generate the vertex degrees. The CG requires a quadratic number of vertices.

All methods we introduced have in common, that they consist of building blocks such as the vertex degree expansion, the zipper scheme or the central processing unit, which consume parts of the neighborhood of the surrounding cluster state. As a result of this space requirement, some graph states cannot be extracted from the two-dimensional cluster state. Finding strong exclusion criteria is challenging, and these criteria may become complex, because there is the possibility that the remaining surrounding cluster state, with the appropriate technique, could still provide the required entanglement structure.

One advantage comes from using the zipper scheme to establish edges, which applies equally to all approaches and schemes. The advantage is that the zipper scheme mainly restores the underlying two-dimensional cluster state, which allows us to cross measurement paths. The crossing of measurement paths enables the extraction of non-plane graph states or the extraction of multiple graph states from a single cluster state.

The centralized approach enables the compact generation of the graph state in a well-defined spatial position, eliminating the need for extensive measurement placement optimization. However, this simplicity comes at the cost of flexibility, as it limits the ability to allow for crossings in the measurement paths and may exclude vertices within the measurement area from the final graph state. Additionally, the generated graph state has to be connected to the intended vertices, leading to further measurements. In contrast, the decentralized approach integrates state generation with the connection process, reducing overall measurement overhead. Though this increases the complexity of measurement path optimization, it allows for greater flexibility in responding to varying requests, as it relies on specific measurement lines rather than large measurement areas.

\section{Discussion}\label{sec:discussion}
In this work, we have introduced a centralized and decentralized approach to extract arbitrary multipartite graph states from a given two-dimensional cluster state by local unitaries and Pauli measurements only. Within the latter approach, we found two schemes, the local vertex degree expansion and the optimized vertex degree expansion, which ensure that the desired degree is reached for each vertex. The centralized approach uses MBQC to generate the desired graph state and then distribute it to the actual vertices.

While we discussed the cluster state sizes from which certain graph states can be extracted, determining the minimal size of a cluster state remains an open question. A related question was answered in Ref.~\cite{dejong2023extracting}, which examined the extraction of GHZ states from linear cluster states. Moreover, it would be beneficial to further explore how to identify the most efficiently extractable graph state within the entanglement class of the desired graph state. Additionally, it would be interesting to develop criteria for assessing which graph states are easier to extract from the two-dimensional cluster state than others. The problem of extracting a graph from a larger graph is an NP-complete problem~\cite{DahlbergNPcomplete}. Nevertheless, sufficiently sparse configurations of the target graph should be extractable in almost any case. Additionally, the derivation of no-go theorems, utilizing rank width~\cite{ChoongbumRankWidth2012} and cut-rank~\cite{NGUYEN2020103183}, for extracting graph states from the two-dimensional cluster state may be a promising direction for further research. Similarly, also entanglement measures such as entanglement between any bipartite cut, provide necessary conditions to check if the transformation to a desired target graph state is possible. Finally, a study of how noise influences the performance of our approaches in the spirit of Ref.~\cite{morruiz2023influence} and its influence on the remaining two-dimensional cluster state would be of practical relevance.

Our approaches are well-suited for entanglement-based quantum networks~\cite{Pirker_2018}, where only local operations generate a request from a pre-distributed state. In this setting, our work gives rise to the study of whether using two-dimensional cluster states as network states provides a substantial improvement over other graph states. For example, one could exchange the GHZ network states in Ref.~\cite{Pirker_2019} by the two-dimensional cluster state. The two-dimensional cluster state has the advantage of allowing rerouting of a measurement path if the scheduled measurement path becomes unavailable, resulting in a more resilient network structure. 

After measuring a vertex in the graph state, it is no longer entangled with the remaining graph state. Especially, suppose each vertex corresponds to a specific network device. In that case, it means that this device cannot participate in the remaining network protocol until the initial large graph state is re-established. An interesting related question is how allowing reconnecting the qubits to the quantum state changes the process.
If the graph state is distributed in a quantum network, where each party holds one qubit, one could use existing quantum channels to distribute Bell states and then connect them via controlled-phase operations to the existing state.

\begin{acknowledgments}  
This research was funded in whole or in part by the Austrian Science Fund (FWF) 10.55776/P36010 and project quantA [10.55776/COE1],  by the Deutsche Forschungsgemeinschaft (DFG, German Research Foundation, project numbers 447948357 and 440958198), the Sino-German Center for Research Promotion (Project M-0294), the German Ministry of Education and Research (Project QuKuK, BMBF Grant No. 16KIS1618K), and the Stiftung der Deutschen Wirtschaft.
For open access purposes, the author has applied a CC BY public copyright license to any author accepted manuscript version arising from this submission.
\end{acknowledgments}

\appendix

\section{Iterative vertex degree expansion}\label{app:vertex_iteratively} 

\begin{figure}
    \centering
    \includegraphics[width=\linewidth]{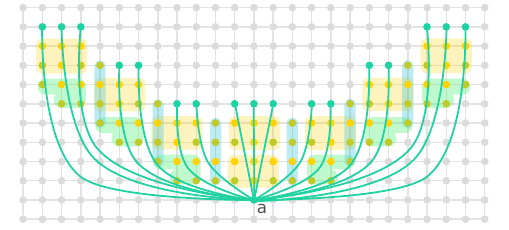}
    \caption{The 'U'-shaped vertex degree expansion increases the degree of the green vertex $a$ located at the bottom center. The green boxes connect one expanded vertex to the adjacent yellow expansion area, resulting in the acquisition of only two vertices through the expansion areas located along the edges of the 'U'. In the corners and the middle of the 'U', the vertex degree increases by three.}
    \label{fig:node_expansion_iterative}
\end{figure}

This Appendix discusses the application of the vertex degree expansion tool from the main text in~\cref{sec:vertex_degree_expansion} along two directions and compares it to the expansion in one direction. The bidirectional application of the vertex degree expansion requires a "U"-shaped area as one sees in~\cref{fig:node_expansion_iterative}. We expand the central, green vertex $a$ at the bottom by applying the vertex expansion (yellow boxes) twice on top. That gives the three top edges in the center and requires $12$ measurements. From the first vertex expansion, we couple out one edge to the right and left, which requires measuring the blue columns and the green patches to transport the lines to the next expansion area. After the edges are coupled out, one can apply the vertex degree expansion again (yellow boxes), which gives us the two green edges on top. Note that the third edge created is used to continue the next expansion area, which again requires the green and yellow measurement patches and one single Pauli $Z$ measurement. Note that we enhance the vertex degree by three at the outer-most and the central expansion because there is no need to get to the next expansion area. If one applies the vertex degree expansion $n_{exp}$ times, $12 \times n_{\text{exp}}+8$ measurements are needed, where the $8$ measurements come from the central out-coupling with the blue columns, and this results in a total expansion by $2\times n_{\text{exp}}$.

We compare the bidirectional expansion with the unidirectional expansion illustrated in~\cref{fig:degree_expansion}\textcolor{magenta}{b} of the main text. Recall that 
applying the vertex degree expansion in one direction $n_{\text{exp}}$ times requires $4\times n_{\text{exp}}$ measurements and enhances the degree by $2\times n_{\text{exp}}$. Therefore, the unidirectional expansion is a factor of three more efficient in terms of required measurements. However, the edges exit on both sides of the measurement area, which may be inconvenient in some cases.

\section{Optimized vertex degree expansion: zipper scheme and merging tool }\label{app:optimized_degree_expansion} 
In this Appendix, we discuss in detail how to combine the merging tool and the zipper scheme to use it for the optimized vertex degree expansion discussed in the main text. In the first part, we investigate how to combine the zipper scheme and the merging tool to connect the desired vertices to a vertex with degree four. In the second part, we investigate the zipper scheme more closely and discuss how it can be adjusted to merge GHZ states with the optimized vertex degree expansion. 

\begin{figure*}
    \centering
    \includegraphics[width=\textwidth]{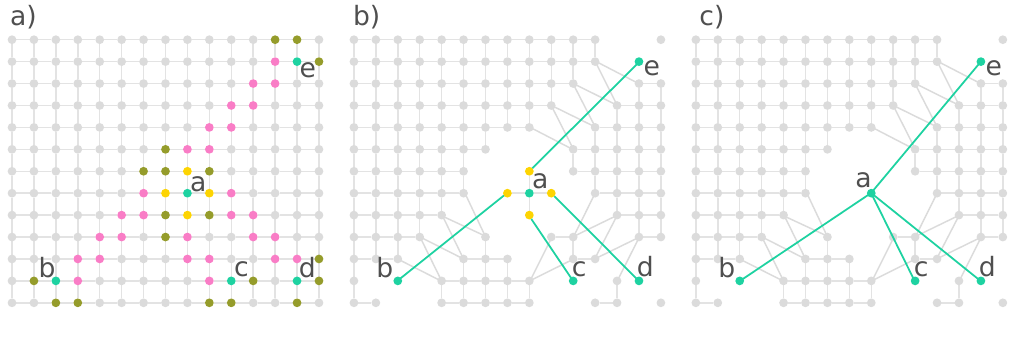}
    \caption{Merging tool and zipper scheme: a) The aim is to connect the vertex with label $a$ to the vertices $b$, $c$, $d$, and $e$. The pink vertices along a staircase-shaped path represent the measurement path for the zipper scheme, and the yellow-green vertices mark the neighbors that need to be removed. b) The yellow vertices around the vertex $a$ are necessary to use the merging 
    tool of the main text. c) The resulting graph state corresponds to a five-qubit GHZ state.}
    \label{fig:GHZ_merging_five}
\end{figure*}

The aim is to directly connect four vertices to an arbitrary vertex that has a vertex degree of four. This can be achieved using the zipper scheme and the merging tool. In~\cref{fig:GHZ_merging_five} we sketch the situation; the aim is to connect the green vertex $a$ to the other green vertices labeled from $b$ to $e$ to obtain the target graph shown in~\cref{fig:GHZ_merging_five}\textcolor{magenta}{c}. In~\cref{fig:GHZ_merging_five}\textcolor{magenta}{a} we show the measurement pattern that allows us to get the desired target state. The first step is to isolate the direct neighbors, yellow vertices, of vertex $a$ by the $Z$ measurements, colored in yellow-green. In this configuration, the yellow vertices can be connected to the desired endpoints $b$ through $e$ using the zipper scheme, indicated by pink vertices aligned along a staircase-shaped path in~\cref{fig:GHZ_merging_five}\textcolor{magenta}{a}. For each junction of the yellow vertices with the zipper scheme, one yellow-green vertex in the Pauli $Z$ basis must be measured. To isolate the vertices $b$ to $e$ Pauli $Z$ measurements on the qubits associated to the vertices colored in yellow-green are needed. \cref{fig:GHZ_merging_five}\textcolor{magenta}{b} shows where the merging tool comes into play, the qubits associated to the yellow vertices are measured in the Pauli $Y$ basis to obtain the target state in~\cref{fig:GHZ_merging_five}\textcolor{magenta}{c}. 

\subsection{Zipper scheme to collect GHZ states} \label{app:local_complementation_scheme}
\begin{figure*}
    \includegraphics[width=\textwidth]{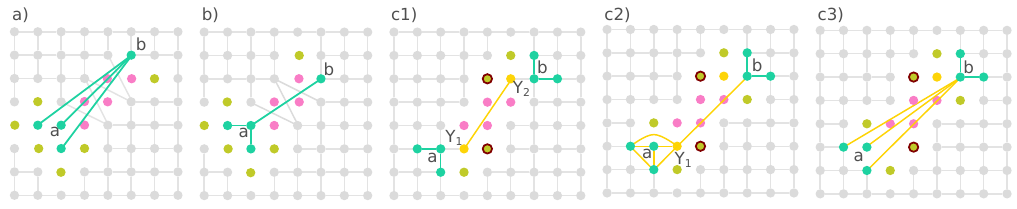}
    \caption{Extracting GHZ states with zipper scheme: a) For an odd number of measurements, the zipper scheme directly allows the extraction of the GHZ state, and the next vertex $b$ turns into the central vertex of the GHZ state. b) The problem occurs for an even number of measurements with the zipper scheme, the vertex $a$ stays the central vertex, and one cannot continue collecting the next vertex using the zipper scheme from $b$. c) Solution for even number of vertices: c1) The first step is to isolate the vertices $Y_1$ and $Y_2$ by the additional Pauli $Z$ measurements in the red circle. c2) Local complementation on $a$ and measuring $Y_2$ in the Pauli $Y$ basis transform the state for the following measurement. c3) Measuring $Y_1$ in the Pauli $Y$ basis transforms vertex $b$ into the central vertex of the GHZ state.}
    \label{fig:collectingGHZ}
\end{figure*}

In the second part of this paragraph, we revisit the zipper scheme in more detail because it helps extracting GHZ states with the optimized vertex degree expansion. The idea is that the central vertex in the GHZ state travels from one vertex to the next using the zipper scheme, thereby collecting the vertices. The number of vertices on the staircase-shaped measurement path influences the measurement strategy. For an odd number of measurements with the zipper scheme, the vertex we gather is already turning into the central vertex, as one sees in~\cref{fig:collectingGHZ}\textcolor{magenta}{a}, and it is ready to collect the upcoming vertices. The situation is different when a staircase-shaped path with an even number of vertices connects the GHZ state; the next vertex does not turn into the central vertex as shown in~\cref{fig:collectingGHZ}\textcolor{magenta}{b}. To overcome that issue, we propose to isolate the first and the last vertex, $Y_1$ and $Y_2$, on the measurement path, which requires measuring an additional neighbor to the path (red circles) for each vertex. Then, the zipper scheme connects the two additional vertices, $Y_1$ and $Y_2$, as one sees in the first sketch in ~\cref{fig:collectingGHZ}\textcolor{magenta}{c1}. Then, one measures the vertex $Y_2$ in the Pauli $Y$ basis and applies local complementation on the vertex $a$ such that locally, a fully connected GHZ state appears, see yellow state in~\cref{fig:collectingGHZ}\textcolor{magenta}{c2}. Finally, we measure the vertex $Y_1$ in the Pauli $Y$ basis, which then turns the vertex $b$ into the root of the GHZ state, as shown in ~\cref{fig:collectingGHZ}\textcolor{magenta}{c3}, from which other vertices can be collected.

\subsection{One turn zipper scheme algorithm}\label{app:one_turn_zipper_algorithm}
We briefly describe our algorithm in~\cref{alg:one_turn_zipper}, which connects two points $a$ and $b$ in a two-dimensional cluster state via staircase-shaped measurement paths with at most a single turn. We determine the intersection coordinate $x_i$ by intersecting two linear functions that go through $a$ and $b$, respectively. Note that two turning points are possible, one above and one below the line that directly connects $a$ and $b$.

\begin{algorithm}
\caption{One turn zipper scheme}
\label{alg:one_turn_zipper}
\KwIn{Vertices to connect $a$ and $b$, cluster state measurement pattern $P$}
\KwOut{Updated cluster state measurement pattern $P$ }
 \tcc{intersection coordinates from ascending staircase trough $a$ and descending staircase trough $b$}
    $x_i = \frac{1}{2}(a_x-a_y+b_x+b_y)$ ;\\
    \If{staircase up to $x_i$ \and and down to $b$ is free  }{
        update $P$ ;
    }
    \Else{
    \tcc{intersection coordinates from descending staircase trough $a$ and ascending staircase trough $b$}
    $x_i = \frac{1}{2}(a_x+a_y+b_x-b_y)$ ;\\
    \If{staircase down to $x_i$ \and and up to $b$ is free  }{
        update $P$ ;
    }
    }
\end{algorithm}

\subsection{Algorithm to find available space in the two-dimensional cluster state  }
In the main text, we mention a second strategy to optimize the vertex expansion, which is using freely available space or strategically clever places in the two-dimensional cluster state for the vertex expansion. In~\cref{alg:free_patches}, we propose a way to find all patches of free vertices $l$ in a two-dimensional grid by iterating through the entire grid $V$. If we find a free vertex $v$, we investigate its environment with the Breadth-First-Search algorithm~\cite{Zuse1948,moore1959shortest} until we find the whole patch $p$ of free vertices connected to $v$. We remove the patch $p$ from the set of vertices $V$ that remains to be visited. Apart from the algorithm we state here, one could also investigate methods from picture analysis or feature recognition techniques to find suitable spaces.

\begin{algorithm}
    \caption{Finding free areas in measurement pattern}
    \label{alg:free_patches}
    \KwIn{Cluster state measurement pattern array $P$}
    \KwOut{Location of free patches $l$ }
    $V = P$ \tcp*{vertex set which has to be visited }
    \While{$V \neq \emptyset$}{
        $v = V(0)$\\
        \If{$v$ is free}{
            \tcc{use Breadth-First-Search to find all free vertices adjacent to $v$}
            $p$ = Breadth\_First\_Search($v$)\\
            $V = V\setminus {p}$\\
            $l = l \cup {p}$
        }
    }
    \Return $l$
\end{algorithm}

\clearpage

\bibliography{main}

\end{document}